\documentclass[a4paper, oneside, twocolumn, notitlepage, 10pt]{extarticle_ecoc}
\usepackage{ecoc}

\begin{document}
\selectlanguage{english}    


\title{Preamble Design and Burst-Mode DSP for Upstream Reception of 200G Coherent TDM-PON}%


\author{
    Haide Wang\textsuperscript{(1)}, Ji Zhou\textsuperscript{(1,*)}, Jinyang Yang\textsuperscript{(1)}, Zhiyang Liu\textsuperscript{(1)}, Cheng Li\textsuperscript{(2)}, Weiping Liu\textsuperscript{(1)}, and \\ Changyuan Yu\textsuperscript{(2)}
}

\maketitle                  


\begin{strip}
 \begin{author_descr}

\textsuperscript{(1)} Department of Electronic Engineering, College of Information Science and Technology, Jinan University, Guangzhou 510632, China, \textcolor{blue}{\uline{zhouji@jnu.edu.cn}} 

\textsuperscript{(2)} Department of Electrical and Electronic Engineering, The Hong Kong Polytechnic University, Hong Kong, China

 \end{author_descr}
\end{strip}

\renewcommand\footnotemark{}
\renewcommand\footnoterule{}


\begin{strip}
    \begin{ecoc_abstract}
        Burst-mode DSP based on 10ns preamble is proposed for upstream reception of 200G coherent TDM-PON. The 128-symbol tone preamble is used for SOP, frequency offset, and sampling phase estimation, while the 192-symbol CAZAC preamble is used for frame synchronization and channel estimation. \textcopyright2024 The Author(s)
    \end{ecoc_abstract}
\end{strip}


\section{Introduction}
In 2021, significant progress was made in the International Telecommunication Union Telecommunication (ITU-T) standardization sector to define a higher-speed (HS) passive optical network (PON) with a line rate of 50Gb/s\cite{ITUT202150gigabit}. Recently, 50G PON has gradually matured with the efforts of vendors and operators. Now, it is time to discuss and research beyond 50G PON. ITU-T sets up a Beyond 50G PON discussion group for discussing future PON-related technologies \cite{ITUT2023b50gigabit}. There are two views on the line rate of Beyond 50G PON: 100Gbit/s or 200Gbit/s. Based on the evolution rules of ITU-T standards, the rate of PON is 4 or 5 times higher than that of previous generation PON. Thus, the rate of Beyond 50G PON may be 200Gbit/s\cite{10023536}. For the 200G PON, direct detection (DD) faces a lot of challenges. Coherent detection becomes the potential solution\cite{10403903}.

Owing to statistical multiplexing, time division multiple access (TDMA) has been used to ensure capacity and quantity for the subscribers since the beginning of the PON commercial applications\cite{9743347}. In 50G DD TDM-PON, one burst-mode trans-impedance amplifier (TIA) and burst-mode DSP are required to receive and process the burst upstream signal, which has a short convergence time to improve spectral efficiency\cite{10484954, 10209815}. Similarly, coherent TDM-PON needs four burst-mode TIAs and burst-mode coherent DSP to process the burst upstream polarization-division-multiplexing (PDM) in-phase and quadrature (IQ) signals such as PDM quadrature phase shift keying and 16-quadrature amplitude modulation (16QAM) signals\cite{9296729, 10117057}. The difficulty of the burst-mode coherent DSP is that the state of polarization (SOP) and frequency offset cause estimation errors in the sampling phase, synchronization position, and tap coefficients of equalizer \cite{8513878}. Therefore, except for the sampling phase, synchronization position, and tap coefficients, the burst-mode coherent DSP should quickly estimate the SOP and frequency offset\cite{10042001}. 

In this paper, we design a 10ns preamble (320 symbols) for the burst-mode DSP in 200G (32GBaud PDM-16QAM) coherent TDM-PON. The 128-symbol tone preamble is used for SOP, frequency offset, and sampling phase estimation, and the 192-symbol constant amplitude zero auto-correlation (CAZAC) preamble is used for frame synchronization and channel estimation.

\section{Experimental setups, preamble design, and burst-mode DSP}

\begin{figure}[!t]
    \centering
    \includegraphics[width=\linewidth]{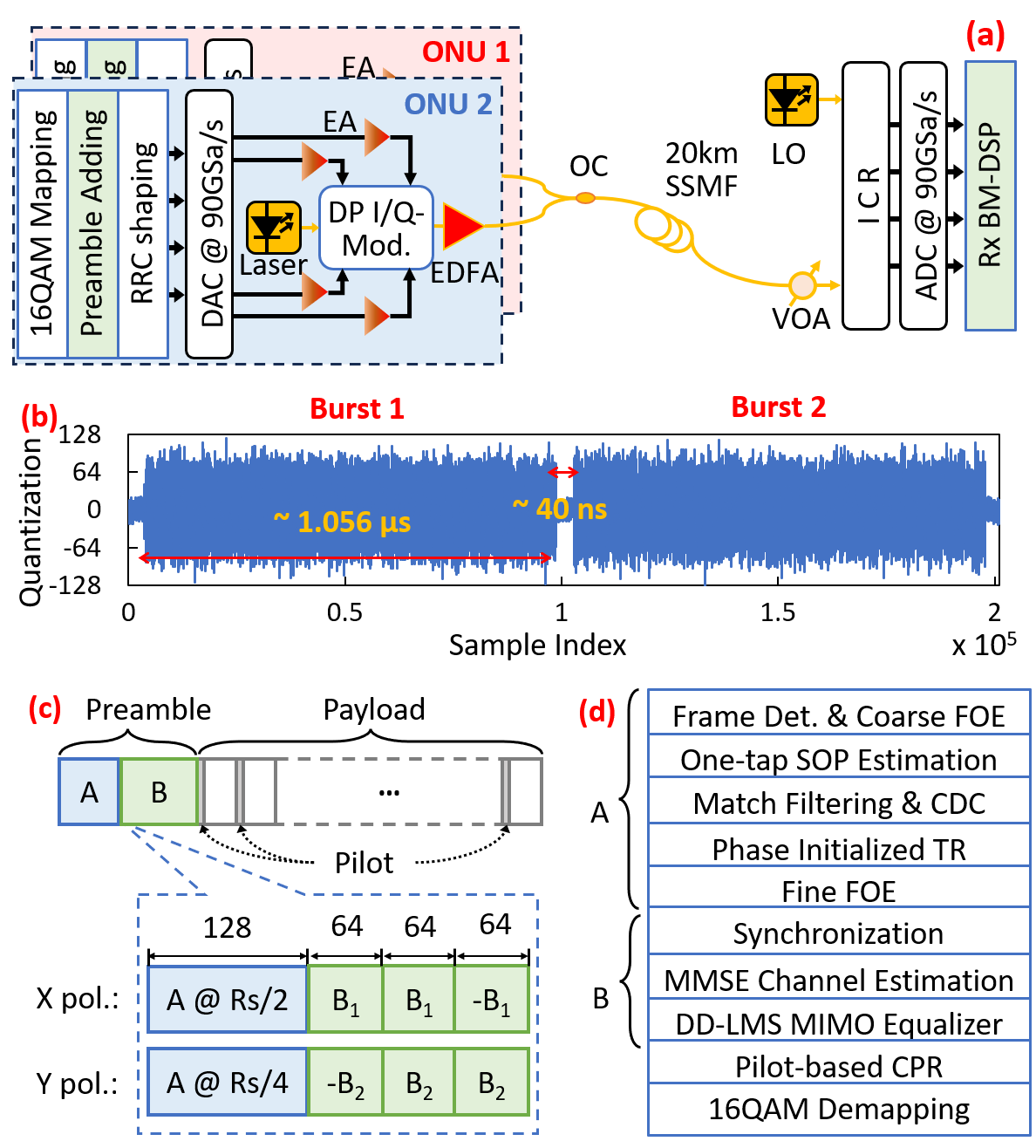}
    \caption{(a) Experimental setups of the 200G coherent TDM-PON in the upstream reception. (b) Two adjacent burst signals. (c) Frame structure and preamble design. (d) Rx Burst-mode DSP (BM-DSP). FF: Feedforward.}
    \label{Ex}
\end{figure}

The experimental setups of the 200G coherent TDM-PON in the upstream reception are shown in Fig. \ref{Ex}(a). At the optical network units (ONUs), a burst signal is generated by the transmitter (Tx) DSP, including 16QAM mapping, preamble adding, and pulse shaping using a raised root cosine (RRC) filter with a roll-off factor of 0.1. Then the digital burst signal is converted to the analog signal by a digital-to-analog converter (DAC) operating at 90GSa/s. After the electrical amplifiers (EAs), the analog signal is modulated on an optical carrier at 1550nm by a dual-polarization in-phase/quadrature modulator (DP I/Q Mod). An external cavity laser (ECL) with a linewidth less than 100kHz is used as the source laser at the Tx. The output power of the modulator is about $-$12.9dBm. An Erbium-doped fiber amplifier (EDFA) boosts the optical power to approximately 3dBm. In commercial scenarios, the semiconductor optical amplifier can be employed. The amplified optical signals of the two ONUs are coupled by a 50:50 optical coupler (OC) and launched to the 20 km standard single-mode fiber (SSMF).

At the optical line terminal (OLT), a variable optical attenuator (VOA) is used to adjust the received optical power (ROP). An integrated coherent receive (ICR) converts an optical signal to an analog signal. An ECL with a power of $\sim$12dBm is employed as a local oscillator (LO). The analog signal is digitized by a 90GSa/s analog-to-digital converter (ADC). Fig. \ref{Ex}(b) shows the two adjacent burst signals. The duration of the burst signal is approximately 1.056 \textmu s, and a guard interval is approximately 40ns. Finally, the signal is recovered by the receiver-side (Rx) burst-mode DSP.

\begin{figure}[t!]
    \centering
    \includegraphics[width=\linewidth]{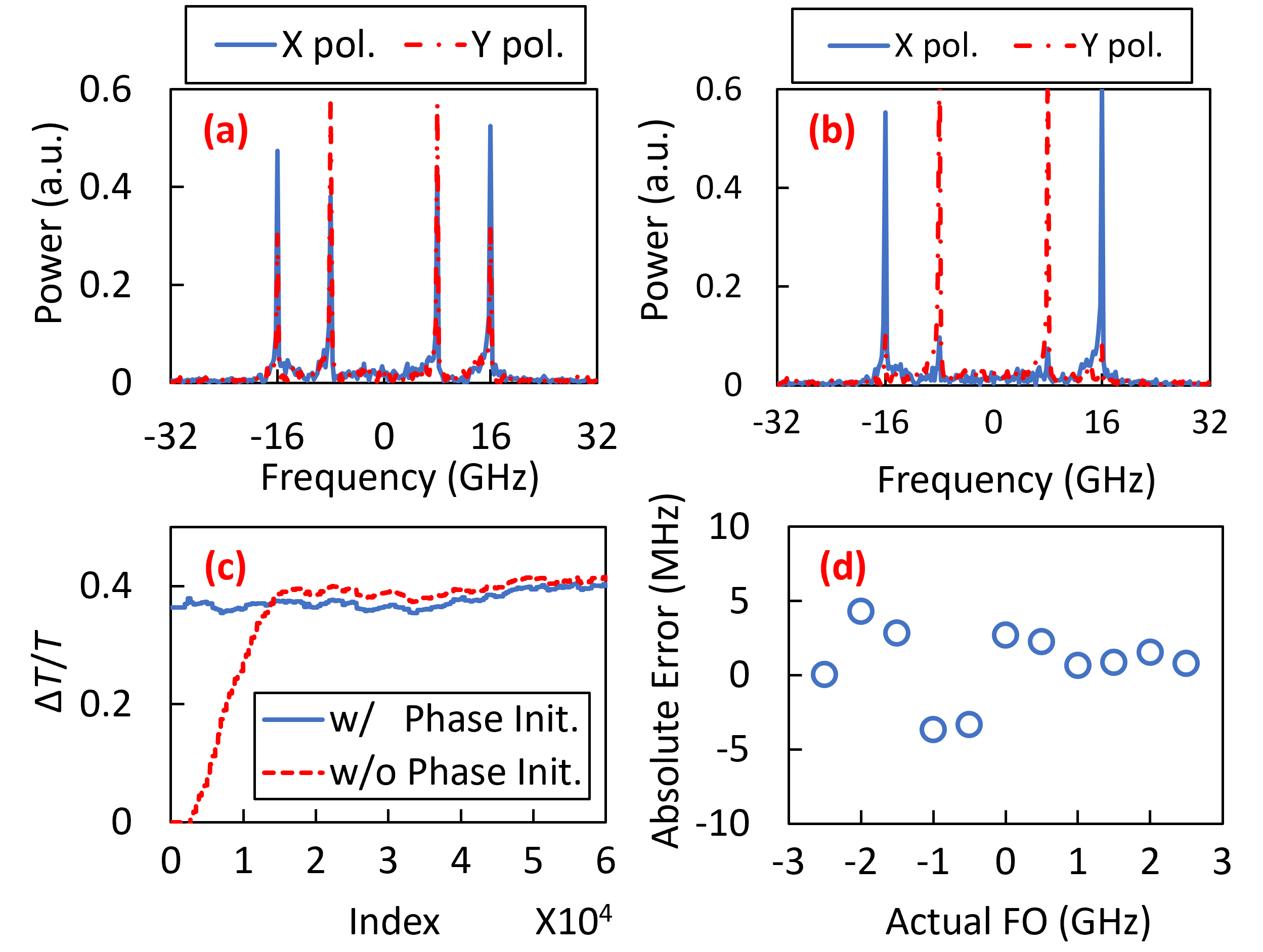}
    \caption{Spectrum of the received Preamble A (a) without and (b) with the SOP estimation and recovery. (c) The sampling phase updated by the TR without and with the sampling phase initialization using Preamble A. (d) Absolute error between the actual frequency offset and estimated frequency offset using the Preamble A.}
    \label{a}
\end{figure}

Fig. \ref{Ex}(c) shows the frame structure and preamble design. The designed preamble consists of two parts, including Preamble A and Preamble B. Preamble A is made up of tone sequences with 128 symbols at half of the baud rate $R_s/2$ and a quarter of the baud rate $R_s/4$ for the X polarization and Y polarization, respectively. Preamble B consists of three 64-symbol CAZAC sequences $\textbf{B}_{1,2}$. For the X polarization, the CAZAC sequence is multiplied by the coefficients of $1$, $1$, and $-1$ for channel estimation and frame synchronization. For the Y polarization, the coefficients are $-1$, $1$, and $1$. One pilot is inserted into every 32 payload for the pilot-based carrier phase recovery (CPR).

Fig. \ref{Ex}(d) shows the Rx burst-mode DSP. Based on Preamble A, frame detection, coarse frequency offset estimation (FOE), and one-tap SOP estimation can be implemented. After the RRC matched filtering with a roll-off factor of 0.1 and chromatic dispersion compensation (CDC), the sampling phase and fine frequency offset are estimated fast by Preamble A to initialize the timing recovery (TR) based on Godard algorithm and implement the fine FOE, respectively. Based on Preamble B, frame synchronization and feedforward channel estimation can be realized. The feedforward channel estimation can be implemented by minimum mean square error (MMSE) and zero-forcing (ZF) algorithms. After initializing the tap coefficients from the channel estimation, the 2 $\times$ 2 multiple-input-multiple-output (MIMO) equalizer is updated by the decision-direct least mean square (DD-LMS) algorithm. After the pilot-based CPR, the recovered 16QAM is demapped into bits.

\section{Experimental results and discussion}
Figs. \ref{a} (a) and (b) show the spectrum of the received Preamble A without and with the SOP estimation and recovery. Before the one-tap SOP recovery, the $R_s/2$ and $R_s/4$ tones crosstalk between the two polarizations. After the one-tap SOP estimation and recovery, most power of the $R_s/2$ tones appears at the X polarization, and most power of the $R_s/4$ tones appears at the Y polarization. Fig. \ref{a} (c) shows the sampling phase updated by the TR with and without the sampling phase initialization using Preamble A. Assuming the loop delay of the TR is approximately 2000 symbols (i.e. 100 symbols/beat$\times$20 beats), the updating based on the Godard algorithm does not work, and the more accurate initialized sampling phase means better performance at the first 2000 symbols. Moreover, the TR with sampling phase initialization can converge faster than that without sampling phase initialization. Fig. \ref{a}(d) shows the absolute error between the actual and estimated frequency offset using the Preamble A. The frequency offset can be calculated accurately with an absolute error within approximately 10MHz.

\begin{figure*}[t!]
    \centering
    \includegraphics[width=\linewidth]{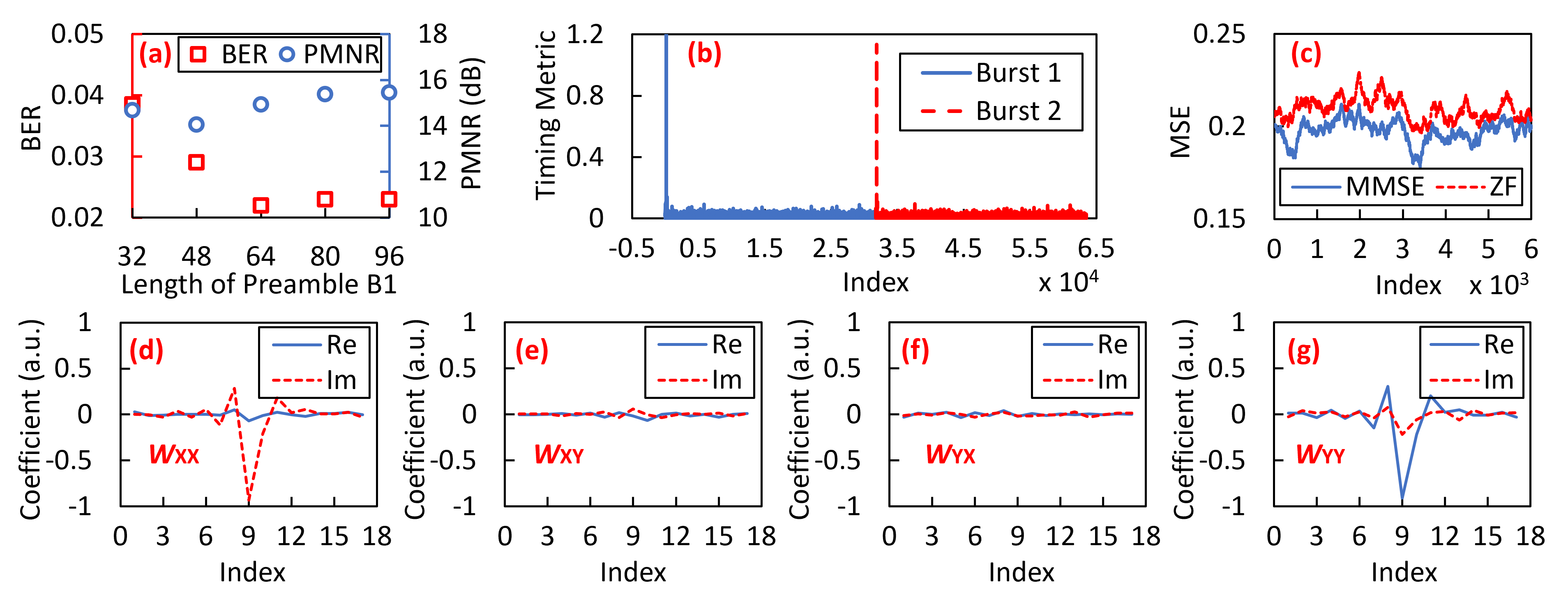}
    \caption{(a) BER of the first one thousand bits and PMNR of the synchronization peak versus the length of preamble $\textbf{B}_1$. (b) Timing metric of the synchronization peaks using the Preamble B for the first and second burst signals. (c) MSE curve of the MIMO equalizer with the initialized tap coefficients estimated by MMSE and ZF algorithms using the Preamble B. (d)-(g) The tap coefficients of the MIMO equalizer estimated by MMSE algorithm using the Preamble B. }
    \label{b}
\end{figure*}

\begin{figure}[t!]
    \centering
    \includegraphics[width=\linewidth]{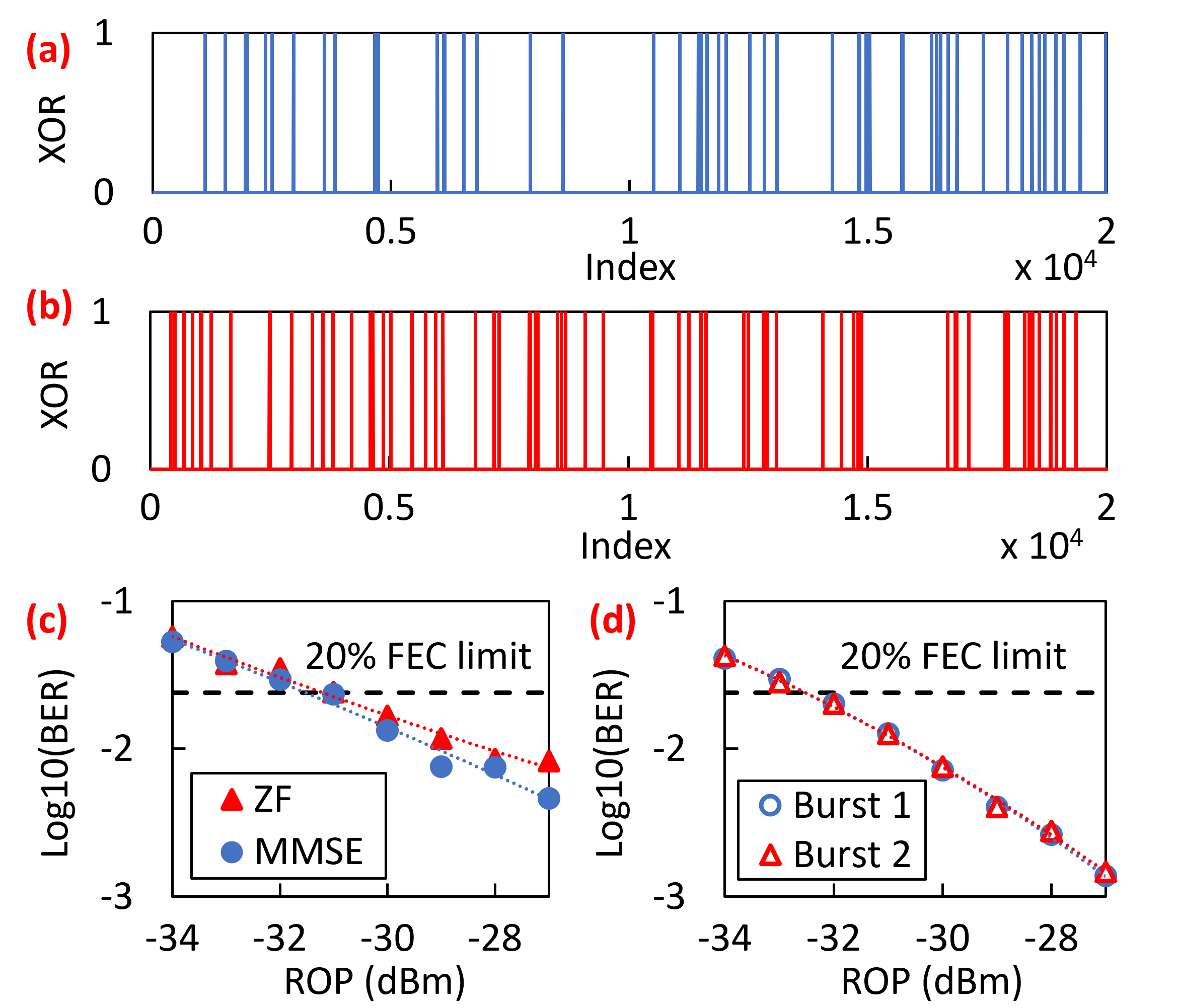}
    \caption{ Bit error distributions using tap coefficients estimated by the (a) MMSE algorithm and (b) ZF algorithm. (c) BER comparison when the MMSE and ZF algorithms are employed. (d) The BER performance of Bursts 1 and 2.}
    \label{BER}
\end{figure}

Fig. \ref{b} (a) shows the bit error ratio (BER) and peak-to-maximum-noise ratio (PMNR) of the synchronization peak versus the length of preamble $\rm{B_1}$ at the ROP of $-$32dBm. The length of the preamble $\textbf{B}_1$ can be set to 64 to achieve a BER under the 20\%-overhead FEC limit of 2.4$\text{e}^{-2}$. Meanwhile, the PMNR is approximately 15dB when the length of the preamble $\textbf{B}_1$ is set to 64. Fig. \ref{b} (b) depicts the timing metric of synchronization. The power of synchronization peaks for Bursts 1 and 2 is much higher than the noise, which means accurate synchronization positions. Fig. \ref{b} (c) shows the mean square error (MSE) curve of the MIMO equalizer with the initialized tap coefficients estimated by MMSE and ZF algorithms using Preamble B. In previous works, the ZF algorithm usually requires a preamble with half zero-padding symbols, which doubles the length of the preamble \cite{10484918}. Owing to the SOP estimation and recovery using Preamble A, the ZF algorithm can be realized using Preamble B under a low computational complexity. However, the MSE of the ZF algorithm is still higher than that of the MMSE algorithm. Fig. \ref{b} (d)-(g) show the tap coefficients $\mathbf{W}_{\text{XX}}$, $\mathbf{W}_{\text{XY}}$, $\mathbf{W}_{\text{YX}}$, and $\mathbf{W}_{\text{XX}}$ for the MIMO equalizer estimated by the MMSE algorithm using Preamble B. The $\mathbf{W}_{\text{XY}}$ and $\mathbf{W}_{\text{YX}}$ are close to 0, which denotes that the SOP has been effectively recovery using the parameters estimated by Preamble A.

Assuming the loop delay of the MIMO equalizer is approximately 6000 symbols (i.e. 100 symbols/beat$\times$60 beats), the updating based on the DD-LMS algorithm does not work, and the more accurate initialized tap coefficients
mean better performance at the first 6000 symbols (i.e. $\sim$20000 bits). Fig. \ref{BER} (a) and (b) show the error distributions of the first 20000 bits using the tap coefficients estimated by the MMSE and ZF algorithms at the ROP of $-$27dBm, respectively. The error distribution using the MMSE algorithm is better than that using the ZF algorithm. Fig. \ref{BER} (c) shows the BER comparison when the MMSE and ZF algorithms are employed. However, at the high ROP, the MMSE algorithm has a clear performance gain compared to the ZF algorithm. At the 20\% FEC limit, MMSE and ZF algorithms have similar performance. Therefore, owing to the SOP estimation and recovery using Preamble A, the low-complexity ZF algorithm based on Preamble B has more potential for estimating the tap coefficients. Fig. \ref{BER} (d) shows the BER performance of Bursts 1 and 2 when the DD-LMS algorithm works. The ROPs of Bursts 1 and 2 are both $-$32dBm. For the uplink of 200G coherent TDM-PON, the optical power budget can achieve 35dB.

\section{Conclusions}
We present the burst-mode DSP based on a 10ns preamble for upstream reception of 200G (32GBaud PDM-16QAM) coherent TDM-PON. The sampling phase of TR and tap coefficients of the MIMO equalizer are estimated accurately and quickly to make the feedback algorithms under the right statuses. It is worth noting that the offline experiments in the Lab are relatively ideal. The preamble length can be properly increased to ensure the performance of the actual scenarios. In conclusion, the proposed burst-mode DSP with a compact preamble makes it more possible to be applied in the future coherent TDM-PON.

\clearpage
\section{Acknowledgements}
This work was supported in part by the National Key R\&D Program of China under Grant 2023YFB2905700, in part by the National Natural Science Foundation of China under Grant 62371207 and Grant 62005102, in part by the Young Elite Scientists Sponsorship Program by CAST under Grant 2023QNRC001, and in part by the Hong Kong Research Grants Council GRF under Grant 15231923.

\defbibnote{myprenote}{
}
\printbibliography[prenote=myprenote]
\vspace{-4mm}

\end{document}